\documentstyle{czjphys}         
\input{epsf}
\def\st{{\scriptscriptstyle T}}
\begin{document}
\title{BOUNDS ON TRANSVERSE MOMENTUM DEPENDENT DISTRIBUTION FUNCTIONS}        
\authori{A. Henneman}      
\addressi{Division of Physics and Astronomy, 
		  Faculty of Science, 
		  Vrije Universiteit\\
		  De Boelelaan 1081,
		  1081 HV Amsterdam, 
		  The Netherlands}     
\authorii{}     
\addressii{}    
\authoriii{}    
\addressiii{}   
\headtitle{BOUNDS ON TRANSVERSE MOMENTUM DEPENDENT DISTRIBUTION FUNCTIONS}            
\headauthor{A. Henneman}           
\specialhead{A. Henneman:BOUNDS ON TRANSVERSE MOMENTUM DEPENDENT DISTRIBUTION FUNCTIONS  \ldots} 
\evidence{A}
\daterec{XXX}    
\cislo{0}  \year{1999}
\setcounter{page}{1}
\pagesfromto{000--000}
\maketitle

\begin{abstract}
When more than one hadron takes part in a hard process, an
extended set of quark distribution and fragmentation functions 
becomes relevant.
In this talk, the derivation of Soffer-like bounds 
for these functions, in the case of a spin-$\frac{1}{2}$ 
target \cite{boundsEigen}, is sketched and some of their 
aspects are discussed.
\end{abstract}

\section{Introduction}
In hard inclusive electro-weak processes the soft physics
is described by light-cone correlators of quark fields.
For a target hadron, for instance, all the relevant soft physics
resides in the correlator \cite{Soper77,Jaffe83,Manohar90}
\begin{equation}
\Phi_{ij}(x) = \left. \int \frac{d\xi^-}{2\pi}\ e^{ip\cdot \xi}
\,\langle P,S\vert \overline \psi_j(0) \psi_i(\xi)
\vert P,S\rangle \right|_{\xi^+ = \xi_\st = 0}.
\end{equation}
Here, $P$ and $S$ denote the parent hadron momentum and spin,
and the relevant component of the quark momentum is $x=p^+/P^+$, 
the light-cone momentum fraction. 
The components $a^\pm = a\cdot n_\mp$ stem from vectors $n_+$ and 
$n_-$, satisfying $n_+^2 = n_-^2$ = 0 and $n_+\cdot n_-$ = 1, which
are fixed by the momentum that introduces the large scale $Q$,
together with a (soft) hadron momentum.
When only the leading part in orders of $1/Q$ is considered, just
the $\Phi \gamma^+$ part of the correlator suffices.
This part is usually parametrized in terms of the following
quark distribution functions~\cite{remark}
\begin{equation}
\Phi(x) \gamma^+ = \Bigl\{
f_1(x) + S_L\,g_1(x)\,\gamma_5  + h_1(x)\,\gamma_5\,{S\!\!\!\slash}_\st
\Bigr\}{\cal P}_+ ,
\label{phiint}
\end{equation}
where ${\cal P}_+$ stands for the projector of good fields 
$\psi_+ \equiv {\cal P}_+\psi = \frac{1}{2}\gamma^-\gamma^+\psi$
\cite{KS70}.
For these functions some trivial bounds and the, less trivial,
so-called Soffer bound have been derived \cite{Soffer73}.

If, now, one regards processes involving more than one
hadron \cite{MT96,TM95}, quark transverse momentum
becomes relevant \cite{RS79}.
A correlator with transverse momentum leads to an extended set 
of distribution functions \cite{kTparam}.
The purpose of this talk is to sketch the derivation of bounds 
for the additional functions in this extended set.
\section{Light-front densities}
%
A first observation in the derivation of the bounds is that
the leading part of the correlator, now being non-diagonal in
target spin space in contrast to (\ref{phiint}),
\begin{equation}
(\Phi\gamma^+)_{ij,s^\prime s} = 
\left. \int \frac{d\xi^-}{2\pi\sqrt{2}}\ e^{ip\cdot \xi}
\,\langle P,s^\prime\vert \psi^\dagger_{+j}(0) \psi_{+i}(\xi)
\vert P,s\rangle \right|_{\xi^+ = \xi_T = 0},
\end{equation}
after inserting a complete set of intermediate states, can be written 
in the following way
\begin{equation}
(\Phi\gamma^+)_{ij,s^\prime s} =
\frac{1}{\sqrt{2}}\sum_n
\langle P_n\vert \psi_{+j}(0)\vert P,s^\prime\rangle^\ast
\langle P_n\vert \psi_{+i}(0)\vert P,s\rangle
\,\delta\left(P_n^+ - (1-x)P^+\right) ,
\label{dens}
\end{equation} 
which is a positive semi-definite quantity.
This property is not affected by inclusion of transverse momentum.

Next, target spin dependence is incorporated using 
a spin density matrix formalism.
\be
M(S) = \mbox{Tr}\left[ \rho(P,S)\,\tilde M(P) \right] 
\label{spindens}
\ee
All target polarization information is in $\rho(P,S)$,
while the spin dependence resides in the higher dimensionality
of $\tilde{M}(P)$.
For a spin-$1/2$ target, $S$ is just a
vector with properties $P\cdot S = 0$ and 
$-1 \le S^2 \le 0$ (being equal to $=-1$ for a pure state) and 
$\tilde{M}$ is just $2 \times 2$ in target spin space.
In the target rest frame $\rho(P,S)$ simplifies to 
$1+{\bf S}\cdot {\bm \sigma}$ and $\tilde{M}$ assumes
the form
\begin{equation}
\tilde M_{ss^\prime} =
\left\lgroup \begin{array}{cc}
M_O + M_L & M_\st^1 - i\,M_\st^2 \\
& \\
M_\st^1 + i\,M_\st^2 & M_O - M_L \\
\end{array}\right\rgroup
\label{spinexplicit}
\end{equation}
where the subscripts refer to target polarization $S=(0,{\bf S}_T,S_L)$,
where $S_L = M S^+/P^+$.
From the diagonal elements of this matrix one sees that it lives in 
the space spanned by states with $S_L = 1$ and $S_L = -1$.
In order to describe transverse target polarization one needs the 
off-diagonal elements.

Now, we turn our attention to quark spin.
The analogon of (\ref{phiint}) when quark transverse momentum is 
taken into account, is given by the sum of three parts
\bea
\Phi_O(x,\bm p_\st)\,\gamma^+ & = &
\Biggl\{
f_1(x,\bm p_\st^2)
+ i\,h_1^\perp(x,\bm p_\st^2)\,\frac{{p \!\!\!\slash}_\st}{M}
\Biggr\} {\cal P}_+
\\
\Phi_L(x,\bm p_\st)\,\gamma^+ & = &
\Biggl\{
S_L\,g_{1L}(x,\bm p_\st^2)\,\gamma_5
+ S_L\,h_{1L}^\perp(x,\bm p_\st^2)
\gamma_5\,\frac{{p\!\!\!\slash}_\st}{M}
\Biggr\} {\cal P}_+
\\
\Phi_T(x,\bm p_\st)\,\gamma^+  & = &
\Biggl\{
f_{1T}^\perp(x,\bm p_\st^2)\,\frac{\epsilon_{\st\,\rho \sigma}
p_\st^\rho S_\st^\sigma}{M}
+ g_{1T}(x,\bm p_\st^2)\,\frac{\bm p_\st\cdot\bm S_\st}{M}
\,\gamma_5
\nonumber \\ & &\mbox{}
+ h_{1T}(x,\bm p_\st^2)\,\gamma_5\,{S\!\!\!\slash}_\st
+ h_{1T}^\perp(x,\bm p_\st^2)\,\frac{\bm p_\st\cdot\bm S_\st}{M}
\,\frac{\gamma_5\,{p\!\!\!\slash}_\st}{M}
\Biggr\} {\cal P}_+.
\eea
Choosing for the above objects a convenient (Weyl) representation, 
one sees that they are effectively $2 \times 2$ in quark spin space.
The leading part of the correlator is spanned by just two types
of quarks; left and right-handed (good) quarks.

If we now put everything together to obtain $\Phi(x,p_T)\gamma^+$
(or it's transpose in dirac space $(\Phi\gamma^+)^T$, to be more 
precise), from an expression like (\ref{spindens}), one concludes 
that the $\tilde{M}$ needed for the description including transverse 
momentum, is the following.
\begin{equation}
\left\lgroup \! \! \! \! 
\begin{array}{cccc} f_1 + g_{1L} & \left( \!\!  \begin{array}{c}
\frac{\vert p_\st\vert}{M}\,e^{i\phi}\times \\
\left(g_{1T}\!+\!i\, f_{1T}^\perp\right)
 \end{array} \! \!  \right)
& \left( \! \!  \begin{array}{c}
\frac{\vert p_\st\vert}{M}\,e^{-i\phi}\times \\
\left(h_{1L}^\perp\!+\!i\,h_1^\perp\right) 
\end{array} \! \!  \right)
& 2\,h_1\\ & & & \\ \left( \! \!  \begin{array}{c}
\frac{\vert p_\st\vert}{M}\,e^{-i\phi}\times \\
\left(g_{1T}\!-\!i\,f_{1T}^\perp\right)
\end{array} \! \!  \right) & f_1 - g_{1L} &
\frac{\vert p_\st\vert^2}{M^2}e^{-2i\phi}\,h_{1T}^\perp &
\left( \! \!  \begin{array}{c} -\frac{\vert p_\st\vert}{M}\,e^{-i\phi}\times \\
\left(h_{1L}^\perp\!-\!i\,h_1^\perp\right) 
\end{array}
\! \!  \right) \\ & & & \\ \left( \! \!  \begin{array}{c}
\frac{\vert p_\st\vert}{M}\,e^{i\phi}\times\\ 
\left(h_{1L}^\perp\!-\!i\,h_1^\perp\right)
\end{array} \! \!  \right) & \frac{\vert p_\st\vert^2}{M^2}e^{2i\phi}\,h_{1T}^\perp &
f_1 - g_{1L} & \left( \! \!  \begin{array}{c} -
\frac{\vert p_\st\vert}{M}\,e^{i\phi}\times \\
\left(g_{1T}\!-\!i\,f_{1T}^\perp\right) 
\end{array} \! \!  \right) \\ & & & \\
2\,h_1 & \left( \! \!  \begin{array}{c}
-\frac{\vert p_\st\vert}{M}\,e^{i\phi}\times \\
\left(h_{1L}^\perp\!+\!i\,h_1^\perp\right) 
\end{array} \! \!  \right) &
\left( \! \!  \begin{array}{c} -\frac{\vert p_\st\vert}{M}\,e^{-i\phi}\times \\
\left(g_{1T}\!+\!i\,f_{1T}^\perp\right) 
\end{array} \! \!  \right) &
f_1 + g_{1L} \end{array} \! \! \!\! \right\rgroup 
\label{bigmatrix}
\end{equation}
This matrix lives in the product space of target helicity and
quark handedness.
The upper-right as the lower-left $2 \times 2$ submatrices are solely
populated by so called chiral-odd functions \cite{Artru,Cortes92,JJ92}, that 
involve the flipping of quark handedness, whereas the diagonal submatrices
contain merely chiral-even functions.
Whithin each of these $2 \times 2$ matrices, the diagonal elements involve 
no and longitudinal polarization, as these states can be expressed in the 
helicity eigenstates of the target, whereas the non-diagonal ones involve 
transverse polarization as expected from (\ref{spindens}).
In (\ref{bigmatrix}) all distribution functions particular to transverse
quark momentum are accompanied by an azimuthal dependence.
This dependence averages to zero after integration over azimuthal angle,
showing that taking into account transverse momentum is necessary to
access the full helicity structure of a polarized nucleon \cite{BoM99}.
The sought for bounds follow from the fact that for any vector $a$ the quantity 
$a \tilde{M} a \ge 0$.
If an integration over azimuthal angle is perfomed first and after that
positive semi-definiteness is demanded, one finds the Soffer bound.
Note that the T-odd functions $f_{1T}^{\perp}$ and 
$h_1^\perp$  can be considered as imaginary parts of 
$g_{1T}$ and $h_{1L}^{\perp}$, respectively.

\section{Interpreting the bounds}
Regarding 2-dimensional subspaces in (\ref{bigmatrix})
starts giving us non-trivial bounds on the distribution functions.
Omitting the $(x,\bm p_\st^2)$ dependences of these functions
one finds
\begin{eqnarray}
&& \vert h_1 \vert \le
\frac{1}{2}\left( f_1 + g_{1L}\right)
\le f_1,
\\
&&
 \frac{{\bm p}_T^2}{2 M^2}\,
\vert h_{1T}^{\perp}\vert \le
\frac{1}{2}\left( f_1 - g_{1L}\right)
\le f_1,
\\
&& 
\frac{{\bm p}_T^4}{4 M^4} \left(
\left( g_{1T}\right)^2
+ \left( f_{1T}^{\perp}\right)^2 \right)
\le \frac{\bm p_\st^2}{4M^2}
\left( f_1 + g_{1L}\right)
\left( f_1 - g_{1L}\right)
\le \frac{\bm p_\st^2}{4M^2}\,f_1^2,
\\
&& 
\frac{{\bm p}_T^4}{4 M^4} \left(
\left( h_{1L}^{\perp}\right)^2
+ \left( h_{1}^{\perp}\right)^2 \right)
\le \frac{\bm p_\st^2}{4M^2}
\left( f_1 + g_{1L}\right)
\left( f_1 - g_{1L}\right)
\le \frac{\bm p_\st^2}{4M^2}\,f_1^2.
\end{eqnarray}
In order to incorporate the more elaborate bounds that are found 
considering higher dimensional subspaces of the matrix (\ref{bigmatrix}),  
it is convenient to introduce two positive-definite functions 
$A(x,\bm p_\st^2)$ and $B(x,\bm p_\st^2)$ such that $f_1 = A + B$ and 
$g_1 = A - B$ and also
\bea
h_1 &=& \alpha\,A ,
\\ 
\frac{{\bm p}_T^2}{2 M^2}h_{1T}^{\perp} &=& \beta\,B ,
\\ 
\frac{{\bm p}_T^2}{2 M^2}
\left( g_{1T} + i\,f_{1T}^{\perp} \right) &=&
\gamma\,\frac{\vert {\bm p}_\st\vert}{M}\,\sqrt{AB} ,
\\ 
\frac{{\bm p}_T^2}{2 M^2} \left(
h_{1L}^{\perp} + i\,h_{1}^{\perp} \right) &=&
\delta\,\frac{\vert {\bm p}_\st\vert}{M}\,\sqrt{AB} .
\eea
Here $\alpha$, $\beta$,$\gamma$ and $\delta$ all depend on both $x$ and 
${\bm p}_T^2$ and have absolute values in the interval $[-1,1]$.
Note that $\alpha$ and $\beta$ are real-valued whereas $\gamma$ and $\delta$ 
are complex-valued.
Their imaginary parts determine the strength of the T-odd functions.

In terms of these functions, what is required to be positive semi-definite
are the following four expressions
\bea
e_{1,2} = (1-\alpha)A + (1+\beta)B
\pm \sqrt{4AB\vert\gamma+\delta\vert^2+((1-\alpha)A-(1+\beta)B)^2},
\\
e_{3,4} = (1+\alpha)A + (1-\beta)B
\pm \sqrt{4AB\vert\gamma-\delta\vert^2+((1+\alpha)A-(1-\beta)B)^2},
\eea
leading to
\bea
&&
A+B\ge 0,
\\ &&
\vert \alpha\,A-\beta\,B \vert \le A+B,
\quad \mbox{i.e.} \ \vert h_{1T}\vert \le f_1,
\\ &&
\vert \gamma + \delta \vert^2 \le (1-\alpha)(1+\beta) ,
\\ &&
\vert \gamma - \delta \vert^2 \le (1+\alpha)(1-\beta) .
\eea

\begin{figure} \centering
\mbox{\epsfysize=40mm\epsffile{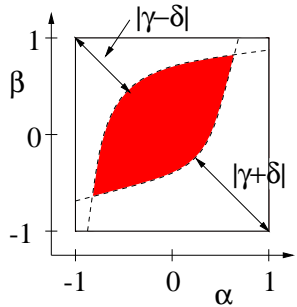}}
\caption{ \label{fig3}
Allowed region (shaded) for $\alpha$ and $\beta$ depending on
$\gamma$ and $\delta$.}
\end{figure}

In figure \ref{fig3} one can see the a graphical representation of
the allowed values for $\alpha$ and $\beta$.
It is remarkable to see that to see that an inclusively measured 
function as $h_1$ is involved in a bound including functions as 
$g_{1T}$ and $h^{\perp}_{1L}$ which cannot be measured 
inclusively and are responsible for asymmetries \cite{MT96,BM98}.
\section{Concluding remarks}
It is important to note that though in this talk only distribution 
functions have been addressed, an almost identical analysis can be
performed on fragmentation functions \cite{CS82}.
The non-vanishing of T-odd functions, though disputed in the case
of distribution functions yet a possibility \cite{Sivers90}, is
accepted in the case of fragmentation functions, such  as 
$D^{\perp}_{1T}$ \cite{MT96,J96} and $H^{\perp}_1$ \cite{Collins93}, 
as time-reversal invariance cannot be imposed on the final state 
\cite{RKR71,HHK83,JJ93}.
This work can also straightforwardly be extended to spin-$1$ hadrons
\cite{alessandro} and gluons \cite{joaoGluons}.
Though it should be supplemented with a study of the
factorization, scheme dependence and stability of the bounds under
$Q^2$ evolution \cite{scheme}, these bounds provide an estimate of 
the magnitudes of functions measured in asymmetries at SMC \cite{SMC}, 
HERMES \cite{HERMES} and LEP \cite{LEP}.


\begin{thebibliography}{30}
\bibitem{remark}
Other common notations for a quark flavor $q$ are $f^q_1(x) = q(x)$, 
$g^q_1(x) = \Delta q(x)$ and $h^q_1(x) = \Delta_T q(x)$.
\end{thebibliography}
\end {document}